\documentclass[aps,prb,amsmath,amssymb,twocolumn,superscriptaddress]{revtex4-1}
\usepackage[pdftex,colorlinks,linkcolor=blue,citecolor=blue,urlcolor=blue]{hyperref}
\usepackage[pdftex]{graphicx}

\newcommand{\FigRef}[2]{\hyperref[#1]{\ref{#1}#2}}
\newcommand{\EqRef}[1]{\hyperref[#1]{(\ref{#1})}}

\begin{document}


\title{
Real spin and pseudospin topologies\\
in the noncentrosymmetric topological nodal-line semimetal CaAgAs
}


\author{Hishiro T. Hirose}
\email{HIROSE.Hishiro@nims.go.jp}
\affiliation{%
  Research Center for Functional Materials,
  National Institute for Materials Science, Tsukuba, Ibaraki 305-0003, Japan
}%
\author{Taichi Terashima}
\affiliation{%
  International Center for Materials Nanoarchitectonics,
  National Institute for Materials Science, Tsukuba, Ibaraki 305-0003, Japan
}%

\author{Taichi Wada}
\affiliation{%
  Department of Applied Physics, Nagoya University, Nagoya 464-8603, Japan
}%

\author{Yoshitaka Matsushita}%
\affiliation{%
  Research Network and Facility Services Division,
  National Institute for Materials Science, Tsukuba, Ibaraki 305-0003, Japan
}%

\author{Yoshihiko Okamoto}
\affiliation{%
  Department of Applied Physics, Nagoya University, Nagoya 464-8603, Japan
}%

\author{Koshi Takenaka}
\affiliation{%
  Department of Applied Physics, Nagoya University, Nagoya 464-8603, Japan
}%

\author{Shinya Uji}%
\affiliation{%
  Research Center for Functional Materials,
  National Institute for Materials Science, Tsukuba, Ibaraki 305-0003, Japan
}%


\date{\today}

\begin{abstract}
  We present the topology of spin-split Fermi surface of CaAgAs
  as determined by de Haas-van Alphen (dHvA) effect measurements
  combined with \textit{ab initio} calculations.
  We have determined the torus-shaped nodal-line Fermi surface from the dHvA oscillations of $\beta$ and $\gamma$ orbits.
  The former orbit encircles the nodal-line, while the latter does not.
  Nevertheless, a nontrivial Berry phase is found for both orbits.
  The nontrivial phase of $\beta$ arises from the orbital characters,
  which can be expressed as a pseudospin rotating around the nodal-line.
  On the other hand, the phase of $\gamma$ is attributed to the vortex of real spin texture
  induced by an antisymmetric spin-orbit interaction.
  Our result demonstrates that both the real- and pseudo-spin textures are indispensable
  in interpreting the electronic topology in noncentrosymmetric nodal-line semimetals.
\end{abstract}



\maketitle


\section{Introduction}

Nodal-line semimetals (NLSMs) are a class of topological materials
characterized by a linearly dispersing band-crossing along a continuous line
in the three-dimensional $k$-space.\cite{A.A.Burkov2011, C.Fang2016}
Various intriguing quantum phenomena are predicted
in NLSM.\cite{N.B.Kopnin2011, 
J.-W.Rhim2015, A.K.Mitchell2015, Y.Huh2016,
L.-K.Lim2017, 
J.P.Carbotte2017, 
S.T.Ramamurthy2017, J.Liu2017, S.P.Mukherjee2017, 
S.V.Syzranov2017, S.Barati2017, 
W.B.Rui2018, 
J.Li2018} 
Although numerous materials are proposed as
the NLSM,
\cite{G.Xu2011, 
G.Bian2013, 
G.Bian2016, 
L.S.Xie2015, 
R.Yu2015, Y.Kim2015, 
L.M.Schoopn2016, M.Neupane2016, 
J.Hu2016, 
D.Takane2016, 
C.Chen2017, 
H.Weng2015, 
K.Mullen2015, 
X.Zhang2017, 
X.Zhang2019, 
Q.F.Liang2016, 
Y.Wu2016, 
S.A.Ekahana2017} 
most candidates accommpany trivial bands around the Fermi level ($E_\text{F}$),
which screen the characteristic properties arising from the nodal-line (NL) bands.
CaAgAs is one of the ideal NLSM which has only a circular NL band around the $E_\text{F}$.
\cite{A.Yamakage2016, E.Emmanouilidou2017, N.Xu2018}

\begin{figure}[b]
  \includegraphics[width=\linewidth]{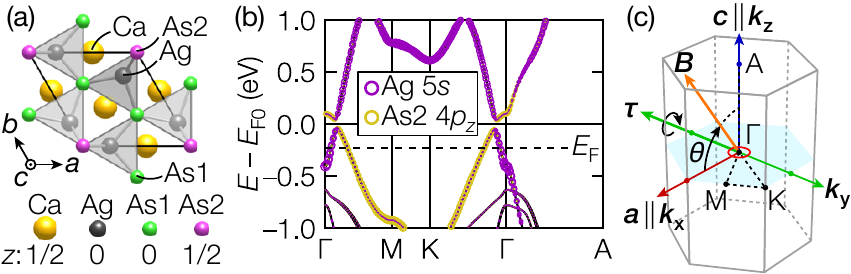}
  \caption{\label{fig:Fig1}
  (Color online)
  (a) Crystal structure of CaAgAs viewed along $c$-axis.
      $z$-parameter of each site is given at the bottom.
  (b) Band structure near the Fermi level.
      Ag $5s$ and As2 $4p_z$ characters are indicated by colors.
      The dashed line indicates the experimental Fermi level $E_\text{F} = -230\,\text{meV}$
      measured from the ideal Fermi level $E_\text{F0}$.
  (c) The red circle around the $\Gamma$ indicates the NL in the Brillouin zone.
      The definitions of $\tau$ and $\theta$ are given.
  }
\end{figure}


CaAgAs crystallizes in the ZrNiAs-type structure with the noncentrosymmetric space group $P\bar{6}2m$ (\#189).\cite{A.Mewis1979}
As depicted in Fig.~\FigRef{fig:Fig1}{(a)},
it consists of four crystallographic sites: Ca, Ag, As1, and As2.
An \textit{ab initio} calculation shows that the conduction and valence bands
mainly consist of Ag $5s$ and As2 $4p_z$ characters, respectively,
which overlap with each other around the $\Gamma$ point
[see Fig.~\FigRef{fig:Fig1}{(b)}].
These orbitals have opposite eigenvalues for the (0001) mirror operation \cite{A.Yamakage2016}
and can be regarded as opposite pseudospins.
Consequently, the bands cannot hybridize at $k_z = 0$ (and $\pi$) without spin-orbit interaction (SOI),
leading to the quarternary degenerated NL
as depicted in Fig.~\FigRef{fig:Fig1}{(c)}.
The perturbation of the SOI allows the hybridization and opens a gap of $\varDelta \sim 75\,\text{meV}$,
giving rise to the strong topological insulator state
for a Fermi energy $(E_\text{F})$ locating in the middle of the gap \cite{C.L.Kane2005, A.Yamakage2016};
though, the NL topology still resides when $E_\text{F}$ is away from the gap.\cite{E.Emmanouilidou2017, C.Li2018}
Experimentally, the linear dispersions associated with the NL bands
are confirmed by angle-resolved photoemission spectroscopy.
\cite{X.-B.Wang2017, J.Nayak2017, D.Takane2018}
However, the effect of spin splitting has not been addressed.
The lack of inversion symmetry lifts the spin degeneracy via an antisymmetric SOI (ASOI),
inducing an additional nontrivial feature of the real spin degree of freedom as in the Rashba and Dresselhaus systems.
\cite{S.-Q.Shen2004, H.Murakawa2013}
Although the ASOI in CaAgAs is small,\cite{A.Yamakage2016} it is still accessible in terms of the quantum oscillation.
Thus, we studied the comprehensive picture of the spin-split Fermi surface (FS) of the NL in CaAgAs.
The nontrivial Berry phase arising from the real spin and that from the pseudospin
are found depending on the trajectory on the torus-shaped FS.

\begin{figure*}
  \includegraphics[width=\linewidth]{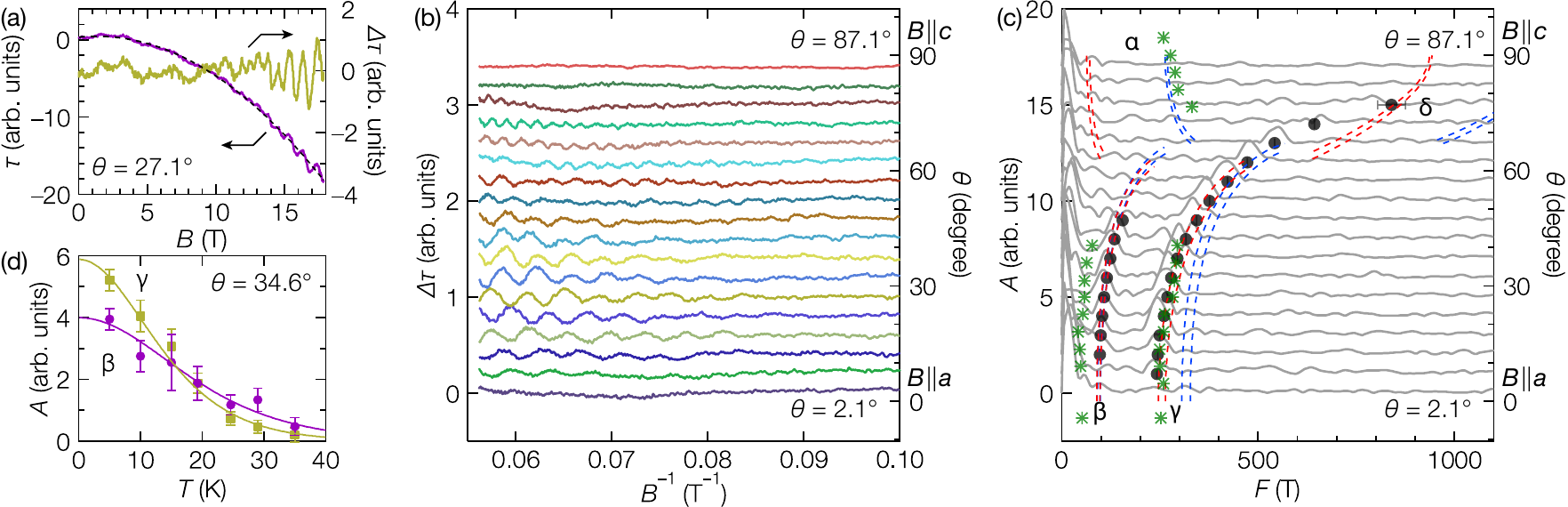}
  \caption{\label{fig:Fig2}
  (Color online)
  (a) $B$ dependence of the $\tau$ (left axis) and $\varDelta\tau$ (right axis) at 30 mK.
      The dashed line indicates a second-order polynomial background.
  (b) $\varDelta\tau$ (left axis) as a function of $B^{-1}$ and
  (c) Fourier transforms of the dHvA oscillations (left axis) for various field directions.
      The data in (b) and (c) are measured with $5^\circ$ interval and
      vertically shifted in accord with $\theta$ (right axis).
      The black circles are average $F$'s of the spin-split FS
      determined by fitting $\varDelta\tau$ at high fields with
      Eq.~\EqRef{eq:individual}.
      The green stars are $F$'s reported in Ref.~\onlinecite{Y.H.Kwan2020}.
      The blue and red dashed lines are $F$ versus $\theta$
      calculated by using the PBE and HSE06 potentials, respectively.
  (d) Temperature dependence of the $\beta$ (circle) and $\gamma$ (square) oscillation amplitudes $A$.
      The errors are defined as a mean background amplitude around the peaks in Fourier transformed spectra
      and the curves are fits to $R_\text{T}$.
  }
\end{figure*}

\section{Methods}
\subsection{Experiments}

Single crystals of CaAgAs were grown as described in Ref.~\onlinecite{D.Takane2018}.
The crystals were confirmed to be a single domain by the X-ray diffraction technique.
The de Haas-van Alphen (dHvA) effect on the magnetic torque $\tau$ was measured with
the piezoresistive cantilever,
\cite{E.Ohmichi2002}
which was rotated in the magnetic field $B$ within the $ac$-plane,
as shown in Fig.~\FigRef{fig:Fig1}{(c)} (see Appendix~\ref{sec:torque_sign} for details).
The field angle $\theta$ is measured from the $a$-axis.

\subsection{Quantum oscillation}

Theoretically, an oscillatory contribution to the magnetic torque
from an extremal orbit $\mathbb{O}$ about the spin-nondegenerate FS
can be described as
\begin{equation}
  \label{eq:individual}
  \varDelta\tau_\mathbb{O} \!=\!
                   CB^{3/2}\frac{\partial F_\mathbb{O}}{\partial\theta}R_TR_\text{D}
                   \sin\!\left[2\pi\!\left(\frac{F_\mathbb{O}}{B} \!-\! \frac{1}{2}\right)
                   \!\pm\!\frac{\pi}{4} \!+\! \phi_\text{Z} \!+\! \phi_\text{B}\right]\!\!,
\end{equation}
where $C$ is a positive coefficient,
$\phi_\text{B}$ is a Berry phase,
and the $\pm$ sign is positive (negative) when $\mathbb{O}$ is a minimum (maximum).\cite{D.Shoenberg1984}
Higher harmonics are neglected.
The frequency $F_\mathbb{O}$ is related to the cross-sectional area $S_\mathbb{O}$ at $B = 0$ of the orbit as
$F_\mathbb{O} = \hbar S_\mathbb{O}/2\pi e$.
The temperature and Dingle reduction factors are given by
$R_T = \xi/\sinh\xi$ and $R_\text{D} = \exp(-\xi_\text{D})$, respectively,
where $\xi_\text{(D)}  = 2\pi^2 k_\text{B}T_\text{(D)}m^*/e\hbar B$,
$T_\text{D}$ is a Dingle temperature, and $m^*$ is a cyclotron effective mass.
The Zeeman energy of electron spin causes
a basically linear-in-$B$ change in the orbit area,
which does not change the apparent frequency of the oscillation
but gives rise to a constant phase shift $\phi_\text{Z}$ expressed as
\begin{equation}
  \label{eq:phi_Z}
  \phi_\text{Z} = \oint_\mathbb{O}
                  \frac{g\hbar\sigma_B}{4m_ev_\perp}|d\mbox{\boldmath{$k$}}|,
\end{equation}
with $\mathbb{O}$ carrying a clockwise orientation.\cite{A.Alexandradinata2018}
Here $g$ is a $g$-factor, $m_e$ is the free electron mass,
and $v_\perp$ is a Fermi velocity along $\mbox{\boldmath{$B$}}\times d\mbox{\boldmath{$k$}}$.
$\sigma_B$ is given by 
$\sigma_B = \hat{\mbox{\boldmath{$B$}}}\cdot\mbox{\boldmath{$P$}}$
with the spin-polarization
$\mbox{\boldmath{$P$}} = \langle\mbox{\boldmath{$\sigma$}}\rangle$.

\subsection{Calculations}

The band-structure, FS, spin polarization and dHvA frequencies ($F$'s) are calculated
from the fully relativistic electronic structure
based on the density functional theory (DFT) \cite{P.Hohenberg1964}
and the tight-binding method.\cite{Wannier90}
More details are given in Appendix~\ref{sec:calculation_method}.
For comparison, we used both the Perdew, Burke, and Ernzerhof (PBE) potential \cite{J.P.Perdew1996} and
the Heyd, Scuseria, and Ernzerhof (HSE06) hybrid potential \cite{G.Kresse1996,J.Heyd2003} in the DFT calculation.

\section{Results}
\subsection{Fermi surface}

Figure \FigRef{fig:Fig2}{(a)} shows $\tau(B)$ at $\theta = 27.1^\circ$,
which is proportional to $B^2$ as expected for paramagnets.
The oscillatory components $\varDelta\tau$ are obtained
by subtracting a second-order polynomial background $\tau_\text{BG}$ from $\tau$,
where dHvA oscillations are discernible above $\sim\!\!10\,\text{T}$.
The angular variation of the $\varDelta\tau$ is plotted against $B^{-1}$ in Fig.~\FigRef{fig:Fig2}{(b)}.
The dHvA oscillations are observed in a wide range of angles ($\theta = 7.1^\circ\text{--}77.1^\circ$).
Figure \FigRef{fig:Fig2}{(c)} shows Fourier transforms of the oscillations
in the range of $9\text{--}17.8\,\text{T}$.
The ASOI-induced spin splitting is too small to be resolved.
We also plot $F$'s determined by fitting the oscillations
at high fields
with Eq.~\EqRef{eq:individual} as circles.
Here, we neglect the spin splitting of the $F$'s
and hence the determined $F$'s are the averages of the split frequencies.
The $F$'s increase as $\theta$ approaches to $90^\circ$.

\begin{figure}
  \includegraphics[width=\linewidth]{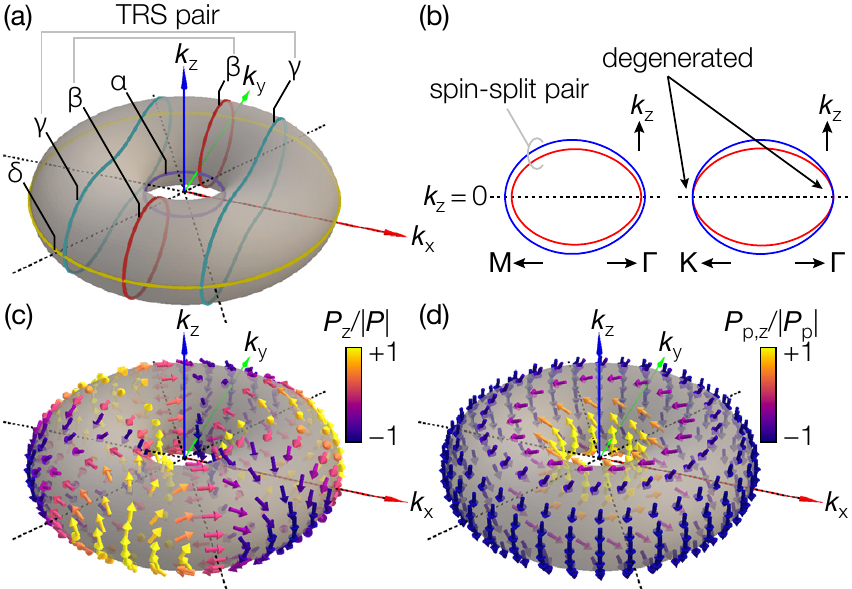}
  \caption{\label{fig:Fig3}
  (Color online)
  (a) Extremal cross-sections of the spin-split FS
      for $B\parallel a$ ($\beta$ and $\gamma$) and $B\parallel c$ ($\alpha$ and $\delta$).
      The dashed lines are along $\Gamma$--K lines.
      The ASOI splits the torus FS into two tori, one nesting inside the other.
  (b) Illustration of the cross-sections of the spin-split FS.
      The red and blue lines indicate the inner and outer tori, respectively.
      The magnitude of spin-split is exaggerated.
  (c) Real spin and (d) pseudospin polarizations for the states of spin-split FS.
      The magnitude of each spin vector is normalized to unity for clarity.
  }
\end{figure}

Figure~\FigRef{fig:Fig3}{(a)}
represents the spin-split FS
calculated with HSE06 potential and $E_\text{F} = -230\,\text{meV}$ (explained below).
The FS of the circular NL becomes torus due to the self-doped hole carriers.\cite{X.-B.Wang2017}
There are four types of extremal orbits:
$\alpha$, $\beta$, $\gamma$, and $\delta$;
the $\alpha$ and $\beta$ ($\gamma$ and $\delta$) orbits correspond to the minimum (maximum) cross-sections.
The ASOI splits the torus into two tori, one nesting inside the other
[Figure~\FigRef{fig:Fig3}{(b)} shows cross-sections schematically].
Accordingly, the four orbits also split into spin-split pairs,
but the splitting is small,
of the order of $1\%$ of the cross-sectional areas.
The Kramers degeneracy is preserved along $\Gamma$-K lines
in consequence of the $D_{3h}$ point-group symmetry.

Figure~\FigRef{fig:Fig2}{(c)} shows the simulated angular dependence of the four frequency-branches
using the PBE and HSE06 potentials with $E_\text{F} = -147,\,-230\,\text{meV}$, respectively,
which are determined so that $F_\beta$ coincides with the experiment.
The smaller (larger) $F$'s correspond to the $\beta$ ($\gamma$ and $\delta$) branch(es).
The overall agreement proves the realization of the torus-shaped FS of the circular NL.
The calculation with the HSE06
gives better agreement with the experiment than the one with the PBE
because the former better estimates the overlap between the conduction and valence bands.

Very recently, Y. H. Kwan \textit{et al.} \cite{Y.H.Kwan2020} reported
measurements of dHvA oscillations on CaAgAs up to $45\,\text{T}$.
They also found the torus-shaped Fermi surface.
The green star marks in Fig.~\FigRef{fig:Fig2}{(c)}
show the $F$'s reported in Ref.~\onlinecite{Y.H.Kwan2020}.
Our $F_\gamma$'s well coincide with theirs
\footnote{Our $\alpha$, $\beta$, $\gamma$, and $\delta$ correspond to
$\beta$, $\delta$, $\gamma$, and $\alpha$ in Ref.~\onlinecite{Y.H.Kwan2020}, respectively.}.
However, our $F_\beta$'s are significantly larger than theirs.
The discrepancy may be ascribed to the difference of frequency resolution
because at most two periods of the $\beta$ oscillation are observed in Ref.~\onlinecite{Y.H.Kwan2020}.
The absence of the $\alpha$ branch in our data is
probably because of the small curvature factor
suppressing its amplitude.\cite{D.Shoenberg1984}
In addition, Y. H. Kwan \textit{et al.} observed
a small peak around $F = 210\,\text{T}$
(not shown in Fig.~\FigRef{fig:Fig2}{[c]}) in their Fourier spectrum
and argued that $F = 210$ and $260\,\text{T}$
might arise from the Zeeman splitting of a single orbit.
However, we saw no corresponding $F$ in our Fourier spectra.
The Zeeman effect does not split $F$'s
but only gives rise to $\pm\phi_\text{Z}$ phase shift as noted above.
We also note that the ASOI-induced spin splitting
($2\varDelta F_\gamma \simeq 3\,\text{T}$ as determined below)
is much smaller than the claimed splitting.

%

Since the oscillation amplitude around $B\parallel a$ is small
due to the small $\partial F_\mathbb{O}/\partial \theta$ factor in Eq.~\EqRef{eq:individual},
we measured $m^*$ at $\theta = 36.4^\circ$.
Figure~\FigRef{fig:Fig2}{(d)} shows
the temperature dependence of the oscillation amplitudes
of $F_\beta(36.4^\circ) = 118\,\text{T}$ and $F_\gamma(36.4^\circ) = 283\,\text{T}$.
$m^*_\beta(36.4^\circ)/m_e = 0.095(9)$, $m^*_\gamma(36.4^\circ)/m_e = 0.130(8)$
are obtained by fitting the data with $R_T$.
Approximating the angular dependence of the $\beta$ orbit as the one of a cylinder along $a$-axis,
we have $F_\beta(0^\circ) \simeq F_\beta(\theta)\cos\theta = 95.0\,\text{T}$ and
$m^*_\beta(0^\circ) \simeq m^*_\beta(\theta)\cos\theta = 0.076(8)\,m_e$,
which correspond to
$k_\text{F} = 5.4\times10^{-2}\,\text{\AA}^{-1}$ and
$v_\text{F} = 8.1(8)\times10^5\,\text{m/s}$ of the $\beta$ cross-section.
Assuming a linear- (parabolic-) dispersion perpendicular to the NL,
the $E_\text{F}$ is estimated as $-288(29)\,[-144(14)]\,\text{meV}$;
the linear-dispersion gives a closer value to $-230\,\text{meV}$
from the \textit{ab initio} calculation, as expected.
The radius of the circular NL $k_\text{R}$ is estimated to be 
$8.4 \times 10^{-2}\,\text{\AA}^{-1}$
from the geometrical relation of the orbits of $F_\beta(36.4^\circ)$ and $F_\gamma(36.4^\circ)$
and assuming an ideally torus-shaped FS.
Accordingly,
the carrier concentration is estimated from the volume of the torus as
$4\pi^2k_\text{R}k_\text{F}^2/(2\pi)^3 = 3.9\times10^{19}\,\text{cm}^{-3}$,
which is smaller than previous reports obtained by the Hall effect.
\cite{E.Emmanouilidou2017, Y.Okamoto2016, D.Takane2018, J.Nayak2017}

Having identified the FS, we visualize,
in Fig.~\FigRef{fig:Fig3}{(c,\,d)},
the calculated polarization of
the real spin $\mbox{\boldmath{$P$}}$ and the pseudospin $\mbox{\boldmath{$P$}}_\text{p}$
on the FS obtained with the HSE06 and $E_\text{F} = -230\,\text{meV}$ determined above.
Here, the up (down) of the pseudospin is defined as the orbital character of
the Ag $5s$ (As2 $4p_z$).
The $\mbox{\boldmath{$P$}}_\text{p}$ is evaluated with the effective eigenspinor constructed
by projecting the calculated tight-binding wavefunction
on the two orbital bases, $|\text{Ag}\,5s\rangle$ and $|\text{As2}\,4p_z\rangle$.
The real spin has a vortex texture around the $\Gamma\text{--K}$ line,
while the pseudospin has one around the NL.

\begin{figure*}
  \includegraphics[width=\textwidth]{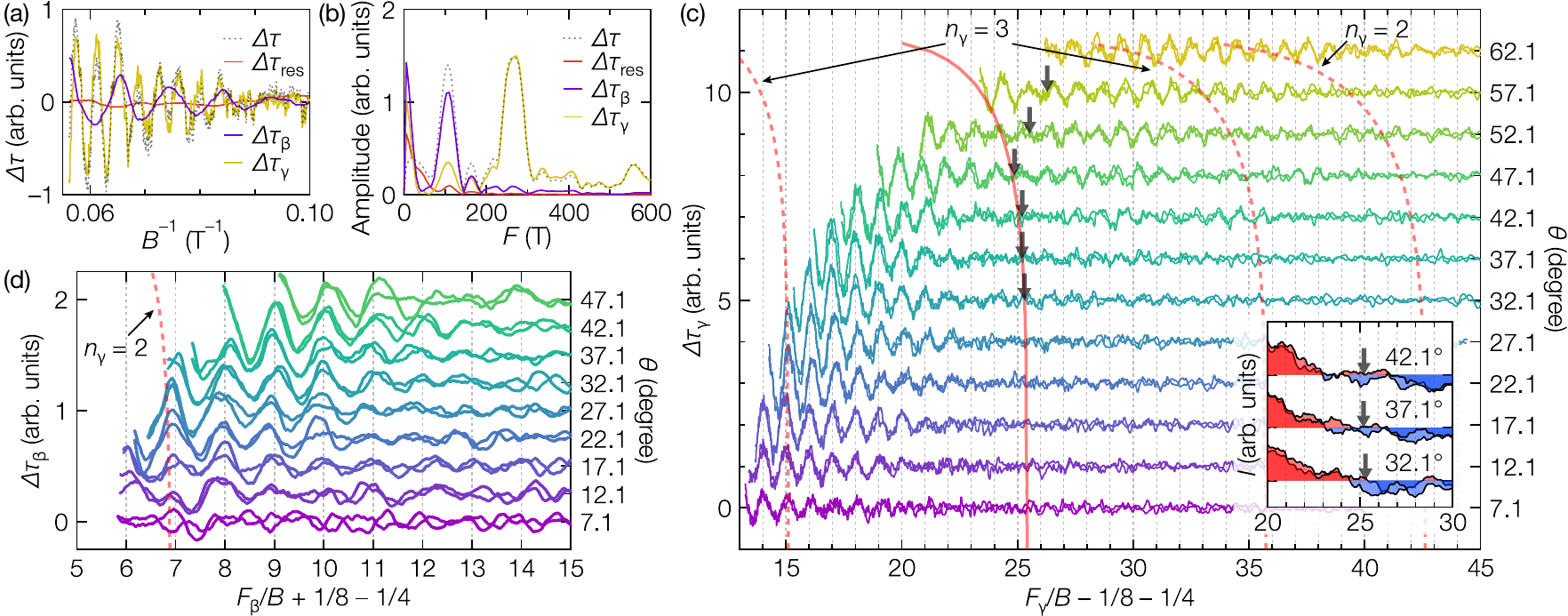}
  \caption{\label{fig:Fig4}
  (Color online)
   (a) Oscillatory components of the $\varDelta\tau$ at $\theta = 27.1^\circ$ plotted against $B^{-1}$.
       $\varDelta\tau_\beta$ and $\varDelta\tau_\gamma$ are extracted from the $\varDelta\tau$ by boxcar smoothing (see text).
       $\varDelta\tau_\text{res}$ is $\varDelta\tau - \varDelta\tau_\beta - \varDelta\tau_\gamma$.
   (b) Fourier transforms of the various $\varDelta\tau$'s shown in (a).
   (c) $\varDelta\tau_\gamma$ and (d) $\varDelta\tau_\beta$
       as a function of $F_{\gamma}/B - 1/8 - 1/4$, and $F_\beta/B + 1/8 - 1/4$, respectively.
       The data are vertically shifted as in Fig.~\FigRef{fig:Fig2}{(b)}.
       To show the reproducibility, the results of two distinct measurements are superimposed.
       The arrows in (c) indicate the positions of the observed beating node.
       The red solid curve is the angle dependence of the node position
       calculated by using $E_\text{ASOI} = 1.37\,\text{meV}$ and the \textit{ab initio} FS,
       while the red dashed curves are the expected neighboring node positions
       if the $n_\gamma$ of the observed node were 2 or 3.
       The inset in (c) shows $I$ defined in Eq.~\EqRef{eq:in-phase}.
  }
\end{figure*}

\subsection{Oscillation phase}

To reveal the nontrivial nature of the electronic states,
we analyzed the phases of the $\beta$ and $\gamma$ oscillations.
In the limit of $B\rightarrow 0$,
neither of the $\beta$ and $\gamma$ orbits is self-constrained
by time-reversal operation;
there is a time-reversal symmetric (TRS) pair of orbits
on each of the spin-split FSs
as indicated in Fig.~\FigRef{fig:Fig3}{(a)}.
Therefore, each of the $\beta$ and $\gamma$ oscillations consists of
the interference of four individual oscillations.

The $\phi_\text{B}$ of an individual oscillation can be considered as
a sum of the real spin contribution $\phi_\text{B,r}$
and the pseudospin contribution $\phi_\text{B,p}$.
Then, the dHvA oscillations from the spin-split pair of orbits
have split frequencies $F_\mathbb{O}\pm\varDelta F_\mathbb{O}$
with the same sign of $\phi_\text{B,p}$
but with opposite signs of $\phi_\text{Z}$ and $\phi_\text{B,r}$.
Similarly, those from the TRS pair of orbits have the same $F$
and opposite signs of $\phi_\text{Z}$, $\phi_\text{B,r}$ and $\phi_\text{B,p}$.\cite{A.Alexandradinata2018}
Thus, the individual oscillations of the $\mathbb{O}\,(= \beta,\,\gamma)$ branch can be expressed as
\begin{align}
  \label{eq:individual_t_s}
  &\varDelta\tau_{\mathbb{O},t,s}=CB^{3/2}\frac{\partial F_\mathbb{O}}{\partial\theta}R_TR_\text{D} \nonumber \\
  &\sin\bigg[2\pi\left(\frac{F_\mathbb{O} + s\varDelta F_\mathbb{O}}{B} - \frac{1}{2}\right)
   \pm\frac{\pi}{4} + ts\phi_\text{Z} + ts\phi_\text{B,r} + t\phi_\text{B,p}\bigg].
\end{align}
Here, $t = \pm 1$ denotes the time-reversal symmetric pair of orbits,
and $s = \pm 1$ denotes the spin-split pair of orbits.
In addition, because the hybridization gap is much smaller than $|E_\text{F}|$,
the $\phi_\text{B,p}$ is constrained to $N\pi$
with $N$ being the winding number of the pseudospin.\cite{G.P.Mikitik1999, C.Li2018, L.Oroszlany2018}
Then, the sum of Eq.~\EqRef{eq:individual_t_s}
for the four individual oscillations of the $\mathbb{O}$ branch becomes
\begin{align}
  \varDelta\tau_\mathbb{O} =\,&\sum_{t,s=\pm 1} \varDelta\tau_{\mathbb{O},t,s}\nonumber \\
                           =\,&4CB^{3/2}\frac{\partial F_\mathbb{O}}{\partial\theta}R_TR_\text{D}
                               \cos\left(2\pi\frac{\varDelta F_\mathbb{O}}{B}\right)
                               \cos\left(\phi_\text{Z} + \phi_\text{B,r}\right)\nonumber\\
                              &\cos\left(\phi_\text{B,p}\right)
                               \sin\left[2\pi\left(\frac{F_\mathbb{O}}{B} - \frac{1}{2}\right)\pm\frac{\pi}{4}\right].
  \label{eq:interference}
\end{align}
The first cosine factor describes
the beating between the spin-split $F$'s,
while the other cosine factors change sign
depending on $\phi_\text{Z}$, $\phi_\text{B,r}$, and $\phi_\text{B,p}$.
In the following, we determine $\phi_\text{B,r}$ and $\phi_\text{B,p}$
for each of $\beta$ and $\gamma$ based on Eq.~\EqRef{eq:interference}. 

To consider the $\beta$ and $\gamma$ oscillations
($\varDelta\tau_\beta$, $\varDelta\tau_\gamma$) separately,
we extract each of them from the observed oscillation $\varDelta\tau$ as follows:
We first plot the $\varDelta\tau$ as a function of $B^{-1}$ in Fig.~\FigRef{fig:Fig4}{(a)}. 
Then, the $\beta$ and $\gamma$ oscillations are effectively suppressed
by applying two boxcar smoothings with the box width of $F_\beta^{-1}$ and $F_\gamma^{-1}$.
The residual $\varDelta\tau_\text{res}$ contains a background from the cantilever.
The $\varDelta\tau_\beta$ is obtained from $\varDelta\tau - \varDelta\tau_\text{res}$
by similarly applying one boxcar smoothing with the box width of $F_\gamma^{-1}$
to remove the $\gamma$ oscillation.
Finally, $\varDelta\tau - \varDelta\tau_\text{res} - \varDelta\tau_\beta$ provides $\varDelta\tau_{\gamma}$.
The results are also shown in Fig.~\FigRef{fig:Fig4}{(a)}.
The Fourier transformations in Fig.~\FigRef{fig:Fig4}{(b)}
confirms the validity of the extraction.
Figures~\FigRef{fig:Fig4}{(c,\,d)} shows $\varDelta\tau_\gamma$ and $\varDelta\tau_\beta$
as a function of $F/B \pm 1/8 - 1/4$.
The sine factor in Eq.~\EqRef{eq:interference}
becomes minima at integers of this abscissa.

Let us start with the $\gamma$ oscillation.
At $\theta = 32.1^\circ$, the sign of the oscillation changes at the specific field $B_\text{node}$,
indicated by arrows in Fig.~\FigRef{fig:Fig4}{(c)};
the oscillation has tops at integers of the abscissa
on the left of $B_\text{node}$ ($B > B_\text{node}$),
while bottoms on the right ($B < B_\text{node}$).
An in-phase intensity with $\cos\left(2\pi x\right)$ in $\varDelta\tau_\gamma$,
\begin{equation}
  \label{eq:in-phase}
  I(x^\prime) = \int_{x^\prime - 1/2}^{x^\prime + 1/2} \varDelta\tau_\gamma(x)\cos\left(2\pi x\right)dx,
\end{equation}
where $x = F_\gamma/B - 1/8 - 1/4$,
shows the sign change across $B_\text{node}$ more apparent
[see the inset of Fig.~\FigRef{fig:Fig4}{(c)}].
This sign change corresponds to the beating due to
the $\cos\left(2\pi\varDelta F_\gamma/B\right)$ factor in Eq.~\EqRef{eq:interference}.
The $B_\text{node}$ is determined by fitting the $\varDelta\tau_\gamma$
with Eq.~\EqRef{eq:interference}.
Note that, at $\theta \ge 47.1^\circ$, there is a finite amplitude of oscillation at $B_\text{node}$
as well as a phase shift and an increase of $F_\gamma/B_\text{node}$ with $\theta$.
They may be explained by an appearance of a magnetic breakdown (MB)
between the spin-split orbits.
A discussion over the MB as well as fittings of $\varDelta\tau_\gamma$ with and without the MB
is given in Appendix~\ref{sec:magnetic_breakdown}.
The $B_\text{node}$ could not be determined for $\theta \le 27.1^\circ$
because the oscillation becomes too weak before the node occurs.

Numbering beating nodes from the highest field one
to satisfy $\varDelta F_\mathbb{O}/B_\text{node} = n_\mathbb{O}/2 - 1/4$ ($n_\mathbb{O} = 1,\,2,\,\ldots$),
the observed one is of $n_\gamma = 1$ or 2.
This is because there is only one beating node within the observed oscillations
ranging from 8\,T to 17.8\,T at each $\theta$.
If $n_\gamma > 2$, the neighboring node $n_\gamma^\prime = n_\gamma \pm 1$ should be observed
at $B_\text{node}^\prime = [2 - (2n_\gamma^\prime + 1)/(2n_\gamma + 1)]B_\text{node}$;
however, no such node exists
[see red dashed curves in Fig.~\FigRef{fig:Fig4}{(c)},
which show expected neighboring node positions when $n_\gamma$ were 2 or 3].


The geometrical relation between the $\gamma$ and $\beta$ orbits
further reduces the possibility of the $n_\gamma$.
If $n_\gamma = 1\,(2)$,
$\varDelta F_\gamma = 2.72\text{--}3.40\,(8.16\text{--}10.20)\,\text{T}$
for $\theta = 32.1\text{--}47.1^\circ$.
Assuming a $k$-independent energy of ASOI $E_\text{ASOI}$,
this corresponds to $E_\text{ASOI} = 1.37\,(4.12)\,\text{meV}$.
Then, the splitting of the $\beta$ oscillation ranges
$\varDelta F_\beta = 1.26\text{--}1.64\,(3.78\text{--}4.93)\,\text{T}$
and the associated position of the beating node for $n_\beta = 1$ is estimated as
$F_\beta/B_\text{node} = 20.8\text{--}20.3\,(6.93\text{--}6.77)$.
The dashed curve in Fig.~\FigRef{fig:Fig4}{(d)}
shows the expected $n_\beta = 1$ node positions when $n_\gamma$ were 2.
The $\beta$ oscillation neither shows a node nor is damped
near the dashed curve at $\theta = 17.1\text{--}32.1^\circ$
where the oscillations are strong enough,
indicating $n_\gamma = 1$ ($F_\beta/B \sim 20.6$ is out of our observation of the dHvA oscillations).

The so determined $n_\gamma = 1$ allows us to find the Berry phase of the $\gamma$ oscillation
from Eq.~\EqRef{eq:interference} as follows:
The $\partial F_\gamma/\partial\theta$ factor is positive (Fig.~\FigRef{fig:Fig2}{[c]}).
Since $\varDelta F_\gamma/B_\text{node} = n_\gamma/2 - 1/4 = 1/4$ at the node position,
$\cos\left(2\pi\varDelta F_\gamma/B\right)$ factor is positive for $B > B_\text{node}$.
While the sine factor of Eq.~\EqRef{eq:interference} takes minima
at integer values of the abscissa as noted above,
the observed $\gamma$ oscillation shows maxima there.
Accordingly, the product $\cos\left(\phi_\text{Z} + \phi_\text{B,r}\right)\cos\left(\phi_\text{B,p}\right)$ is negative.
Moreover, since the $\gamma$ orbit is self-constrained by the $(01\bar{1}0)$ mirror operation
as long as $B$ is rotated within the $k_x\text{--}k_z$ plane,
the $\phi_\text{B,r}$ is constrained to an integer-multiple of $\pi$.
\cite{[{This constraint is mentioned in Ref.~\onlinecite{A.Alexandradinata2018} as the class II-A ($u = 1$, $s = 0$)}]Comment1}
From the same reason, $\phi_\text{Z}$ is always 0 as given in Appendix~\ref{sec:phi_Z}.
Therefore, the $\gamma$ orbit has a nontrivial Berry phase arising from either of $\phi_\text{B,r}$ or $\phi_\text{B,p}$.
Since the $\gamma$ orbit topologically does not encircles the NL,
$\phi_\text{B,p} = 0$ and $\phi_\text{B,r} = \pi$ $(\mathrm{mod}\,2\pi)$ are concluded.
This result agrees with the expectation from the fact that
the $\gamma$ orbit encircles three $\Gamma\text{--K}$ lines, leading to $\phi_\text{B,r} = 3\pi$. 
Thus, the nontrivial Berry phase of the $\gamma$ orbit is attributed to the real spin texture.

Similarly, the Berry phase of the $\beta$ oscillation is determined:
From Fig.~\FigRef{fig:Fig2}{(c)}, $\partial F_\beta/\partial \theta > 0$.
As mentioned above,
the $\beta$ oscillations are observed
in the field range $B > B_\text{node}$ for $n_\beta = 1$;
hence, $\cos\left(2\pi\varDelta F_\beta/B\right) > 0$.
Since the $\beta$ oscillation shows maxima at integer values of the abscissa,
the $\cos\left(\phi_\text{Z} + \phi_\text{B,r}\right)\cos\left(\phi_\text{B,p}\right)$ factor
is identified as negative for the angle range $\theta = 17.1\text{--}47.1^\circ$,
where we observe the discernible $\beta$ oscillation.
In the case of the $\beta$ orbit,
the constraint on the $\phi_\text{B,r}$ \cite{Comment1} (Appendix~\ref{sec:phi_Br}) and
$\phi_\text{Z} = 0$ (Appendix~\ref{sec:phi_Z}) are assured only at $\theta = 0^\circ$
where the orbit is self-constrained by the $(0001)$ mirror operation.
However, it can be shown from the elaborate spin-zero analysis
given in Appendix~\ref{sec:spin-zero} that
the sign of $\cos\left(\phi_\text{Z} + \phi_\text{B,r}\right)$ factor
does not change in $|\theta| \le 47.1^\circ$.
Consequently, the $\beta$ orbit also has a nontrivial Berry phase at $\theta = 0^\circ$
owing to either of the $\phi_\text{B,r}$ or $\phi_\text{B,p}$.
Contrary to the $\gamma$ orbit, the $\beta$ orbit encircles no $\Gamma\text{--K}$ line but encircles the NL.
Therefore, $\phi_\text{B,p} = \pi$, which is attributed to the NL
and evidences the NL topology of the orbital characters.

Finally, it should be mentioned that
Y. H. Kwan \textit{et al.} reached a conflicting conclusion in Ref.~\onlinecite{Y.H.Kwan2020}:
They concluded that the $\beta$ orbit has a nontirvial Berry phase $\pi$,
whereas the $\gamma$ orbit does not.
Ref.~\onlinecite{Y.H.Kwan2020} used a different analytical procedure than we do,
but that is not the source of the discrepancy as follows:
Ref.~\onlinecite{Y.H.Kwan2020} determined the Berry phases of the $\beta$ and $\gamma$ oscillations
by fitting the total oscillation with the simplified two-component Landau-Lifshitz formula,
neglecting the ASOI-induced spin splitting of the Fermi surface.
Although this type of analysis could be superior in that
it could allow one to investigate the exact phase values,
the spin splitting cannot be neglected
since Berry phases obtained from analysis depend on
the sign of the $\cos\left(2\pi\varDelta F_\mathbb{O}/B\right)$ beating factor in Eq.~\EqRef{eq:interference}
where $2\varDelta F_\mathbb{O}$ is the spin splitting at $B = 0$.
According to our analysis,
the beating nodes observed in our $\gamma$ oscillation
(arrows in Fig.~\FigRef{fig:Fig3}{[c]}) are the highest-field ones
and thus $\cos\left(2\pi\varDelta F_\mathbb{O}/B\right) > 0$
at higher fields as implicitly assumed in Ref.~\onlinecite{Y.H.Kwan2020}.
Therefore, the neglect of the beating factor in Ref.~\onlinecite{Y.H.Kwan2020}
cannot explain the discrepancy.
In passing, the fact that Ref.~\onlinecite{Y.H.Kwan2020} observed
no sign of a beating node at higher fields
proves the correctness of our analysis.

Although the exact reason of the discrepancy is unclear,
we can point out the following possible factors:
(1) The sign of the torque signal used for the analysis in Ref.~\onlinecite{Y.H.Kwan2020}
was not determined experimentally,
but assumed to be negative based on that CaAgAs is diamagnetic.
However, the sign of the torque is determined by the anisotropy of the susceptibility
and depends on the field direction (see Appendix~\ref{sec:torque_sign}).
Therefore it is unclear
whether the correct sign was assigned to the analyzed torque signal in Ref.~\onlinecite{Y.H.Kwan2020}.
(2) Ref.~\onlinecite{Y.H.Kwan2020} used much higher magnetic fields, $45\,\text{T}$,
where the MB between spin-split orbits may not be negligible.
The MB may affect the phase of the $\gamma$ oscillation
as described in Appendix~\ref{sec:magnetic_breakdown}.
(3) The phase of the $\beta$ oscillation
determined in Ref.~\onlinecite{Y.H.Kwan2020} has considerable ambiguity
since at most two periods of the $\beta$ oscillation are discernible in Ref.~\onlinecite{Y.H.Kwan2020}.
Note that there is a discrepancy in $F_\beta$ values
between theirs and ours (Fig.~\FigRef{fig:Fig2}{[c]}).
(4) It is unclear toward which direction the field was tilted in Ref.~\onlinecite{Y.H.Kwan2020};
the field was rotated from $c$-parallel to ``$c$-perpendicular'' directions.
Since the phase of the $\gamma$ oscillation depends on
the number of encircled $\Gamma\text{--K}$ lines,
as well as constraints by symmetry,
the phase depends on the field direction.

\section{Conclusion}

In conclusion, we have determined the torus-shaped FS in CaAgAs
via quantum-oscillation measurements.
We have analyzed the oscillations
by taking into account the interference of oscillations
from the ASOI-induced spin-split pair of orbits
as well as the TRS pair of orbits.
As a result, we have found a nontrivial Berry phase for both $\beta$ and $\gamma$ orbits.
The former encircles the NL and hence the observed Berry phase is ascribable to the pseudospin texture around the NL.
The latter orbit topologically does not encircle the NL.
With the aid of \textit{ab initio} calculations,
we have demonstrated that the Berry phase associated with $\gamma$ originates from the real spin texture
where the spin direction rotates around the $\Gamma\text{--K}$ line in the Brillouin zone.
Our results suggest that noncentrosymmetric NL semimetals provide
fertile ground for investigating new quantum phenomena arising from synergy
between spin and orbital pseudospin physics.
\appendix

\section{Sign of magnetic torque\label{sec:torque_sign}}
\begin{figure}
  \includegraphics[width=\linewidth]{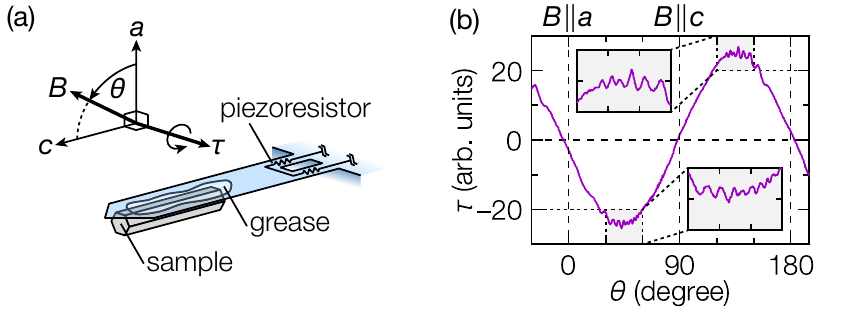}
  \caption{\label{fig:sign_of_torque}
  (a) Schematic of the experimental setup for the torque magnetometry
      utilizing a cantilever and the notations for $\theta$ and $\tau$.
  (b) Angular dependence of the $\tau$ measured at $17.8\,\text{T}$ and $30\,\text{mK}$.
      The insets are the enlarged views of the angular variation of the dHvA oscillaitons.
  }
\end{figure}

We measured the magnetic torque $\tau$ by using a piezoresistive cantilever
(MouldLessCantilever SSI-SS-ML-PRC400, Seiko Instruments Inc.).\cite{E.Ohmichi2002}
The experimental setup is schematically illustrated
in Fig.~\FigRef{fig:sign_of_torque}{(a)}
together with the notations for the field angle $\theta$ and $\tau$.
The sign of $\tau$ exerted on a sample is known
from whether the resistance of the piezoresistor increases or decreases.
The sign of $\tau$
is essential when discussing the phase of de-Haas van-Alphen (dHvA) oscillation;
assigning a wrong sign of the oscillation shifts the phase by $\pi$.
Figure~\FigRef{fig:sign_of_torque}{(b)}
shows the angular dependence of $\tau$ at $17.8\,\text{T}$ and $30\,\text{mK}$.
Since the $\tau$ is expressed in terms of a magnetic susceptibility $\chi$ as
$\tau = -(\partial\chi/\partial\theta)B^2$,
the sign of the sinusoidal torque curve
in Fig.~\FigRef{fig:sign_of_torque}{(b)}
is consistent with the anisotropy of $\chi$, $\chi_a < \chi_c$,
measured on a single crystal;
this confirms the sign of our torque data.
The insets of Fig.~\FigRef{fig:sign_of_torque}{(b)}
show enlarged views of
the $\theta$ variation of dHvA oscillations superimposed on the torque curve.
The signs of the dHvA oscillations
in $0^\circ < \theta < 90^\circ$ and in $90^\circ < \theta < 180^\circ$
are opposite due to the different sign of the $\partial F_\mathbb{O}/\partial \theta$ factor
in Eq.~\EqRef{eq:individual}.

\section{Calculation method}\label{sec:calculation_method}
The fully relativistic electronic structure was calculated based on density functional theory \cite{P.Hohenberg1964}
as implemented in the Quantum ESPRESSO package.\cite{QE}
For comparison, we used both the Perdew, Burke, and Ernzerhof (PBE) function \cite{J.P.Perdew1996} and
the Heyd, Scuseria, and Ernzerhof (HSE06) hybrid function \cite{G.Kresse1996,J.Heyd2003} for exchange potential.
A $6\times6\times9$ $k$-point mesh was used for
the self-consistent field procedure.
A plane-wave cutoff energy of 140 Ry
and a fully relativistic projector augmented-wave method \cite{A.D.Corso2014}
were used for
the calculation with the PBE potential,
while a plane-wave cutoff energy of 55 Ry,
fully relativistic norm-conserving pseudopotentials,
\cite{D.R.Hamann2013,M.Schlipf2015,P.Scherpelz2016}
and a $2\times2\times3$ $q$-point mesh were used for
the calculation with the HSE06 potential.
The difference of the cutoff energies is due to the different types of the pseudopotentials.
The band-structure, Fermi surface, and spin polarizations are calculated
by using the 54-orbital tight-binding model based on maximally localized Wannier functions
constructed with the Wannier90 program.\cite{Wannier90}
The dHvA frequencies are calculated from the Fermi surface
by using the algorithm described in Ref.~\onlinecite{SKEAF}.


\begin{figure*}
  \includegraphics[width=\textwidth]{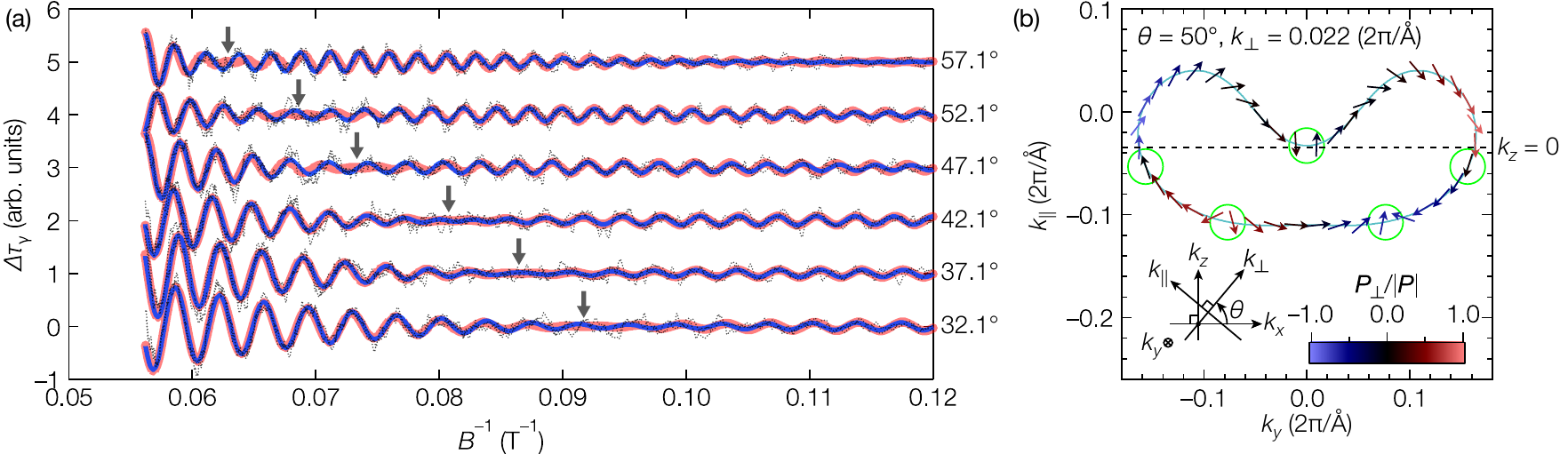}
  \caption{\label{fig:B_node_fit}
  (a) $\gamma$ oscillation component $\varDelta\gamma$ as a function of $B^{-1}$.
      The gray dotted lines are experimental data,
      whereas the bold red and blue curves are the fits with Eq.~\EqRef{eq:interference}
      and Eq.~\EqRef{eq:total}, respectively.
      Arrows indicate the positions of $B = B_\text{node}$.
  (b) Real spin polarization $\mbox{\boldmath{$P$}}$ on the $\gamma$ orbit at $\theta = 50^\circ$.
      The $\gamma$ orbit is on $k_\parallel\text{--}k_y$ plane at $k_\perp = 0.022\,(2\pi/\text{\AA})$,
      where $k_\parallel = -\sin(50^\circ)k_x + \cos(50^\circ)k_z$ and
      $k_\perp = \cos(50^\circ)k_x + \sin(50^\circ)k_z$.
      The inset illustrates the relation between the $k_\parallel\text{--}k_\perp$ and $k_x\text{--}k_z$ coordinates.
      The direction and color of the arrows indicate
      the in-plane and out-of-plane components of $\mbox{\boldmath{$P$}}/|\mbox{\boldmath{$P$}}|$, respectively.
      Green circles mark the candidates of the breakdown $k$ points where $\mbox{\boldmath{$P$}}$ rotates quickly along the orbit.
  }
\end{figure*}

\section{Magnetic breakdown\label{sec:magnetic_breakdown}}

The interference of four individual oscillations results in
the beating with the envelope function $\cos(2\pi\varDelta F_\mathbb{O}/B)$.
We observe such a beating in the $\gamma$ oscillation.
The magnetic field at the beating node $B_\text{node}$
is obtained by fitting the $\varDelta\tau_\gamma$ with Eq.~\EqRef{eq:interference}.
Here, we omit $\partial F_\gamma/\partial\theta$, $\cos(\phi_\text{Z} + \phi_\text{B,r})$,
and $\cos(\phi_\text{B,p})$ factors since they are only related to the intensity and the sign.
The so obtained $B_\text{node}$s are indicated by arrows in Fig.~\FigRef{fig:B_node_fit}{(a)}
together with the red fitting curves.

It is noticeable in Fig.~\FigRef{fig:B_node_fit}{(a)}
that $\varDelta\tau_\gamma$s for $\theta \ge 47.1^\circ$ have a finite intensity of oscillation
even at $B = B_\text{node}$.
Since $\varDelta F_\gamma/F_\gamma \sim 0.011$ is quite small,
the difference of $\partial F_\gamma/\partial \theta$ factor, the effective mass, or the Dingle temperature
between the spin-split orbits may not account for the intensity at $B_\text{node}$.
A MB between the spin-split orbits is rather plausible origin
because the intensity at $B_\text{node}$ becomes larger as $B_\text{node}$ increases with $\theta$.
MB is an electron tunneling between two distinct extremal orbits at specific $k$ points.
When an electron completes a closed orbit with an even number of MBs,
it contributes to the dHvA oscillation whose frequency $F_\text{MB}$
corresponds to the area enclosed by its trajectory;
$F_\text{MB}$ is between $F_\gamma\pm\varDelta F_\gamma$.
Generally, the MB between spinless bands can occur
when the cyclotron energy $\hbar\omega_\text{c} = \hbar eB_\text{c}/m^*$ exceeds $E_\text{g}^2/E_\text{F}$,
where $E_\text{g}$ is an energy gap between the orbits and $E_\text{F}$ is the Fermi energy.\cite{D.Shoenberg1984}
By using $E_\text{F} = 288\,\text{meV}$ and $m^*_\gamma = 0.130\,m_e$ at $\theta = 36.4^\circ$ and
approximating $E_\text{g}$ as $2E_\text{ASOI} = 2.75\,\text{meV}$ obtained in the main text,
$B_\text{c}$ is estimated as $\sim 0.03\,\text{T}$.
The quite small $B_\text{c}$ indicates that the MB can occur
when the spin polarization can be neglected.

In case of the MB between the spin-split bands,
tunneling between the opposite spin state is expected to be suppressed.\cite{N.Kimura2018}
This would be also the case of CaAgAs, where the energy scale of the spin-orbit interaction (SOI) $\varDelta \sim 75\,\text{meV}$
is far larger than $\hbar eB/m^* \sim 16\,\text{meV}$ at $B = 17.8\,\text{T}$.
An exception is at $k$ points where spin orientation quickly rotates to the opposite along the orbit;
an electron tunnels so that to preserve the spin orientation.
The spin polarization on the $\gamma$ orbit for $\theta = 50^\circ$ is shown
in Fig.~\FigRef{fig:B_node_fit}{(b)}.
There are five $k$ points, indicated by circles, where spin polarization quickly changes;
those are candidates of the breakdown $k$ points where MB may occur.
The observed MB oscillation is probably a sum of several MB oscillations
corresponding to the MBs occurring at any possible selection of the breakdown $k$ points.
The intensity of the MB oscillation at $B = B_\text{node}$ decreases with decreasing $\theta$
and almost vanishes at $\theta = 42.1^\circ$.
This trend may indicate that the MB only occurs at $B > B_\text{c}\sim 13\,\text{T}$.
The discrepancy between the $B_\text{c}$s estimated from
the spinless assumption and the intensity at $B = B_\text{node}$ is
probably because an electron needs to tunnel much longer distance (and larger $E_\text{g}$)
than the spinless case to preserve the spin orientation.

The phase shift and the increase of $F_\gamma/B_\text{node}$ observed
in $\varDelta\tau_\gamma$ at $\theta \ge 47.1^\circ$
can be also explained by considering the effect of MB.
The effect of MB can be introduced into Eq.~\EqRef{eq:interference}
as an additional factor $R_{m,n} = (ip)^{m}(q)^{n}$,
where $p^2 + q^2 = 1$, $p^2$ ($q^2$) is the probability of (not) having MB at the breakdown $k$ point,
and $m$ ($n$) is the number of MBs (not) taking place at a breakdown $k$ points in an orbit.
Assuming that the probabilities of having a MB at each breakdown $k$ points are equivalent,
it is expressed as $p^2 = \exp(-B_\text{c}/B)$.
Since our data is not sufficient to decompose the MB oscillations to each,
we roughly approximate the MB oscillation as a single component
which has a factor of $\alpha R_{2, 3}$,
a frequency of $F_\text{MB} = F_\gamma$,
and an arbitrary phase shift $\phi_\text{MB}$.
The $\alpha$ is a correction factor to take into account contributions from all MB oscillations.
The phase shift occurs because the electron does not complete its orbit in a single band.
Then, the MB oscillation for the $\gamma$ orbit is expressed as
\begin{align}
  \label{eq:MB}
  \varDelta\tau_\text{MB} \simeq\,&\,4\alpha CB^{3/2}\frac{\partial F_\gamma}{\partial\theta}R_TR_\text{D}R_{2,3} \nonumber \\
                           &\sin\left[2\pi\left(\frac{F_\gamma}{B}
                            - \frac{1}{2}\right)
                            - \frac{\pi}{4} + \phi_\text{MB}\right].
\end{align}
By taking a sum with the non-MB oscillation $\varDelta\tau_\gamma$ multiplied by $R_{0,5}$,
the total oscillation becomes
\begin{align}
  \label{eq:total}
  &\varDelta\tau_\text{total} = 4CB^{3/2}\frac{\partial F_\gamma}{\partial\theta}R_TR_\text{D} \nonumber \\
  &\hspace{40pt}\left(X^2 + Y^2\right)^{1/2}
   \sin\!\left[2\pi\left(\frac{F_\gamma}{B} - \frac{1}{2}\right) - \frac{\pi}{4} + \phi_\text{MB}^\prime\right],
\end{align}
where
\begin{align}
  &X = R_{0,5}\cos\!\left(2\pi\frac{\varDelta F_\gamma}{B}\right)
       \cos\!\left(\phi_\text{Z} + \phi_\text{B,r}\right)\cos\!\left(\phi_\text{B,p}\right) \nonumber \\
  &\hspace{20pt} + \alpha R_{2,3}\cos\left(\phi_\text{MB}\right), \nonumber \\
  &Y = \alpha R_{2,3}\sin\left(\phi_\text{MB}\right), \nonumber \\
  &\sin(\phi_\text{MB}^\prime) = Y/\sqrt{X^2 + Y^2}, \nonumber \\
  &\cos(\phi_\text{MB}^\prime) = X/\sqrt{X^2 + Y^2}. \nonumber
\end{align}
The $\phi_\text{MB}^\prime$ explains the observed phase shift.
The node position of the envelope function corresponds to the minimum of $(X^2 + Y^2)^{1/2}$,
where $B_\text{node}$ no longer satisfies $\varDelta F_\gamma/B_\text{node} = n_\gamma/2 - 1/4$
due to the non-zero $R_{2,3}$ factor.
Thus, the increase of $F_\gamma/B_\text{node}$ at $\theta \ge 47.1^\circ$ may stem from the MB.
The Eq.~\EqRef{eq:total} well reproduces the observed $\varDelta\tau_\gamma$,
as shown in Fig.~\FigRef{fig:B_node_fit}{(a)}.
Since the effect of MB is not apparent at $\theta \le 42.1^\circ$,
our analyses and results based on $B_\text{node}$ are not affected by MB.

\section{Constraints on $\mbox{\boldmath{$\phi_\text{Z}$}}$\label{sec:phi_Z}}
The value of $\phi_\text{Z}$ can be deduced from the $D_{3h}$ point-group symmetry
and the symmetry of an orbit.
For the ease of understanding,
we give a parametric representation of the spin-polarization
$\mbox{\boldmath{$P$}}(\mbox{\boldmath{$k$}})$ 
up to third order of $k$ \cite{P.A.Frigeri2005}:
\begin{equation}
  \label{eq:spin-texture}
  \mbox{\boldmath{$P$}}(\mbox{\boldmath{$k$}})
     = \alpha_1k_r^2k_z(\hat{\mbox{\boldmath{$P$}}}_x\sin2\phi
     + \hat{\mbox{\boldmath{$P$}}}_y\cos2\phi)
     + \alpha_2k_r^3\hat{\mbox{\boldmath{$P$}}}_z\sin3\phi.
\end{equation}
Here, $\alpha_1$ and $\alpha_2$ are independent coefficients,
$k_r = (k_x^2 + k_y^2)^{1/2}$, and $\phi = \arctan(k_y/k_x)$.
This form well reproduces the real spin texture
from the \textit{ab initio} calculation shown in Fig.~\FigRef{fig:Fig3}{(c)}.

In the case of the $\gamma$ orbit,
the orbit is self-constrained by the $(01\bar{1}0)$ mirror operation
since $B$ is rotated within the $(01\bar{1}0)$ mirror plane
(which is equivalent to $k_x\text{--}k_z$ plane and $a\text{--}c$ plane).
Then, for any $\mbox{\boldmath{$k$}} = (k_x, k_y, k_z)$ on the $\gamma$ orbit,
$\mbox{\boldmath{$k$}}^\prime = (k_x, -k_y, k_z)$ exists on the same orbit and
$P_i(\mbox{\boldmath{$k$}}) = -P_i(\mbox{\boldmath{$k$}}^\prime)\,(i = x,\,z)$
according to Eq.~\EqRef{eq:spin-texture}.
Thus, $\sigma_\text{B} = \mbox{\boldmath{$\hat{B}$}}\cdot\mbox{\boldmath{$P$}}$
in Eq.~\EqRef{eq:phi_Z} cancels out within the orbit, leading to $\phi_\text{Z} = 0$.
When $B$ is strong enough to align $\sigma$ along $B$, the cancellation is not valid.
However, the energy scale of the SOI is $\varDelta \sim 75\,\text{meV}$,
which is far larger than the Zeeman energy of $\sim\!\!1\,\text{meV}$ at 17.8 T and $g = 2$.
So, the cancellation is valid.

In the case of the $\beta$ orbit, the situation is similar
at $\theta = 0^\circ$ ($B \parallel k_x)$.
The $\beta$ orbit is self-constrained by the (0001) mirror operation at $\theta = 0^\circ$.
Then, for any $\mbox{\boldmath{$k$}} = (k_x, k_y, k_z)$ on the $\beta$ orbit,
$\mbox{\boldmath{$k$}}^\prime = (k_x, k_y, -k_z)$ exists on the same orbit and
$P_i(\mbox{\boldmath{$k$}}) = -P_i(\mbox{\boldmath{$k$}}^\prime)\,(i = x,\,y)$
according to Eq.~\EqRef{eq:spin-texture}.
Therefore, $\sigma_\text{B}$ cancels out within the orbit.
This can also be confirmed simply because
the $\beta$ orbit at $\theta = 0^\circ$ locates on the $k_x = 0$ plane,
where $P_x(\mbox{\boldmath{$k$}})$ is restricted to 0 due to the $D_{3h}$ point-group symmetry.
As a result, the $\sigma_B$ in Eq.~\EqRef{eq:phi_Z} is 0, and hence $\phi_\text{Z} = 0$.
On the other hand, at $\theta > 0^\circ$,
$\mbox{\boldmath{$B$}}$ is no longer perpendicular to $\mbox{\boldmath{$P$}}$.
Thus, the increase of the $\sigma_B$ is proportional to $\sin\theta$
by approximating the $\mbox{\boldmath{$P$}}(k)$ as being parallel to $k_z$.
Besides, $\phi_\text{Z}$ is proportional not only to $\sigma_B$ but also to $m^*_\beta$
since $v_\perp = \hbar k_\perp/m^*$.
By approximating the $\theta$ variation of the $\beta$ orbit
as the one of a cylinder along the $k_x$-axis,
$m^*_\beta(\theta)$ is expressed as $m^*_\beta(0^\circ)/\cos\theta$.
Therefore, the $\phi_\text{Z}$ of the $\beta$ roughly increases as $\propto\tan\theta$.

\section{Real spin Berry phase of the $\mbox{\boldmath{$\beta$}}$ orbit\label{sec:phi_Br}}
\begin{figure}
  \includegraphics[width=\linewidth]{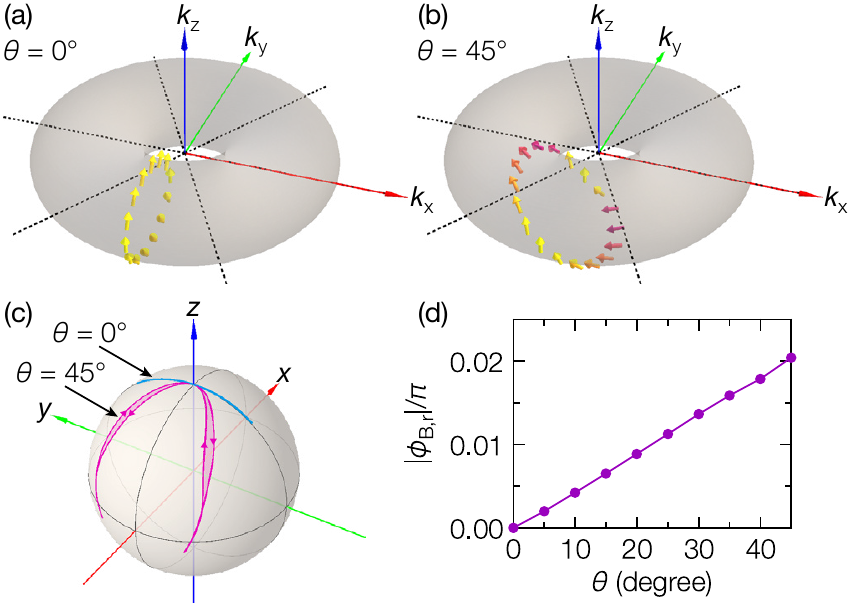}
  \caption{\label{fig:Spin_Berry_beta}
  (a,\,b)
      Real spin polarization $\mbox{\boldmath{$P$}}$ on the $\beta$ orbit at (a) $\theta = 0^\circ$ and (b) $45^\circ$.
  (c) The trajectories of $\mbox{\boldmath{$P$}}$ along the $\beta$ orbit
      at $\theta = 0^\circ$ and $45^\circ$ projected on the Bloch sphere.
  (d) Angular dependence of the $|\phi_\text{B,r}|$ for the $\beta$ orbit.
  }
\end{figure}

As mentioned in the main text,
the $\phi_\text{B,r}$ of the $\beta$ orbit
is constrained to an integer-multiple of $\pi$ only at $\theta = 0^\circ$ by the $(0001)$ mirror operation,
whereas it deviates from the constrained value at $\theta > 0^\circ$.
Here we show how the $\phi_\text{B,r}$ is constrained at $\theta = 0^\circ$
and how small the deviation of the $\phi_\text{B,r}$ is at $\theta > 0^\circ$
based on the \textit{ab initio} calculation.

Figure~\FigRef{fig:Spin_Berry_beta}{(a)} shows
the $\mbox{\boldmath{$P$}}$ at $k$ points on the $\beta$ orbit
at $\theta = 0^\circ$.
The $\mbox{\boldmath{$P$}}$ is restricted within the $k_y\text{--}k_z$ plane
due to the $D_{3h}$ point-group symmetry,
as mentioned in Appendix~\ref{sec:phi_Z}.
Consequently, the trajectory of the $\mbox{\boldmath{$P$}}$ along the $\beta$ orbit projected on the Bloch sphere
sweeps out zero solid angle,
as shown in Fig.~\FigRef{fig:Spin_Berry_beta}{(c)}.
As this solid angle directly corresponds to the twice of the Berry phase,\cite{D.Vanderbilt2018}
the $\phi_\text{B,r}$ of the $\beta$ orbit at $\theta = 0^\circ$ is zero.

In contrast, at $\theta = 45^\circ$, 
the $\mbox{\boldmath{$P$}}$ on the $\beta$ orbit
shown in Fig.~\FigRef{fig:Spin_Berry_beta}{(b)}
is not restricted within the $k_y\text{--}k_z$ plane.
Hence, the projected trajectory shown in Fig.~\FigRef{fig:Spin_Berry_beta}{(c)}
is deformed from the arc of $\theta = 0^\circ$.
However, the solid angle swept out by the trajectory is quite limited,
and the corresponding $\phi_\text{B,r}$ is as small as $0.02\,\pi$.
This is because the $\beta$ orbit locates within the local $k$-space
where $\mbox{\boldmath{$P$}}(k)$ is a slowly varying function of $k$, away from the vortex structure.
Besides, the angular dependence of the $\phi_\text{B,r}$ represented in
Fig.~\FigRef{fig:Spin_Berry_beta}{(d)}
shows that the $|\phi_\text{B,r}|$ monotonically increases from 0
as $\theta$ varies from $0^\circ$.
Therefore, neglecting the angular dependence of the $\phi_\text{B,r}$
when analyzing the experimental data
does not affect the result.

\section{Spin-zero analysis on the $\mbox{\boldmath{$\beta$}}$ oscillation\label{sec:spin-zero}}

In the $\beta$ orbit,
the angular variation of the $\phi_\text{B,r}$ is negligibly small (see Appendix~\ref{sec:phi_Br}),
whereas the $\phi_\text{Z}$ increases in proportional to $\tan\theta$ (see Appendix~\ref{sec:phi_Z}).
Since we could not observe an apparent $\beta$ oscillation at $|\theta| < 17.1^\circ$,
it is crucial to determine whether the $\phi_\text{Z}$ changes
the sign of $\cos\left(\phi_\text{Z} + \phi_\text{B,r}\right)$ factor in Eq.~\EqRef{eq:interference} against $\theta$.
This is similar to the spin-zero analysis widely conducted on
the (quasi-) 2D materials with spin-degeneracy (at $B \rightarrow 0$).
\cite{M.K.Kartsovnik2004, S.E.Sebastian2012, T.Terashima2018, Y.Obata2019}
As seen in Fig.~\FigRef{fig:Fig1}{(d)} in the main text,
$\varDelta\tau_\beta$ does not change the sign against $\theta$
between $17.1\text{--}47.1^\circ$.
If there is a sign-change between $0^\circ$ and $17.1^\circ$,
there should be another sign-change between $17.1^\circ$ and $47.1^\circ$
because $\phi_\text{Z} \propto \tan\theta$ grows more rapidly as $\theta$ increases.
This fact indicates that the sign of 
$\cos\left(\phi_\text{Z} + \phi_\text{B,r}\right)$ does not change in $|\theta| \le 47.1^\circ$.

\begin{acknowledgments}
  This work was partially supported by JSPS KAKENHI Grant No. JP17H07349, and No. JP17H01144.
\end{acknowledgments}


\providecommand{\noopsort}[1]{}\providecommand{\singleletter}[1]{#1}%

\end{document}